\documentstyle[11pt,fleqn]{article}

\newcommand{\R}{\mbox{$I\!\!R$}}             
\newcommand{\beq}{\begin{eqnarray}}         
\newcommand{\eeq}{\end{eqnarray}}

\begin{document}


\hfill{\sl preprint - UTF 403 \\ hep-th/9706191 }
\par
\bigskip
\par
\rm


\par
\bigskip
\begin{center}
\bf
\LARGE
$\zeta$-function renormalization of one-loop stress tensors
in curved spacetime, a check on the method  in the conical manifold
and other cases.
\end{center}
\par
\bigskip
\par
\rm
\normalsize



\begin{center}\Large

Valter Moretti \footnote{e-mail:
\sl moretti@science.unitn.it} 

\end{center}



\begin{center}
\small
\smallskip
European Centre for Theoretical Nuclear Physics and Related Areas\\
Villa Tambosi, Strada delle  Tabarelle 286
I-38050 Villazzano (TN), Italy
\\
and
\\
 Dipartimento di Fisica, Universit\`a di Trento,
 Istituto Nazionale di Fisica Nucleare,
Gruppo Collegato di Trento, via Sommarive 14 38050 Povo (TN), Italy

\end{center}\rm\normalsize



\par
\bigskip
\par
\hfill{\sl June 1997}
\par
\medskip
\par\rm



\begin{description}
\item{Abstract:} 
\it\\
A previously introduced method to renormalize the 
one-loop stress tensor and the one-loop vacuum fluctuations 
in a curved background by a direct
local $\zeta$-function approach is checked in some thermal and 
nonthermal cases.\\
First the method is checked in the case of a conformally coupled
massless field in the  static Einstein universe where all hypotheses initially
requested by the method hold true. Secondly, dropping the hypothesis of a 
closed manifold, the method is checked in the open static Einstein universe.
Finally, the method is checked for a massless scalar field  in the presence 
of a conical singularity in the Euclidean manifold (i.e. Rindler 
spacetimes/large mass black hole manifold/cosmic string manifold). 
In all cases, a complete agreement with other approaches is found.\\
Concerning  the last case in particular,
the method is proven to 
give rise to the stress tensor already got by the point-splitting approach 
for every coupling with the curvature regardless of the presence of the 
singular curvature.
In the last case, comments  on the measure employed in the path integral, the
use of the optical manifold and  the different approaches to renormalize 
the Hamiltonian are made.
\par
\end{description}
\rm


\smallskip
\noindent{\sl PACS number(s):\hspace{0.3cm}
04.62.+v}
\par
\bigskip
\rm



\section*{Introduction}

In \cite{Zmoretti} we have introduced a direct  method to compute the 
one-loop Euclidean renormalized stress tensor through a  $\zeta$
function approach. The considered stress tensor is that obtained
by varying the one-loop effective 
action $S_{\scriptsize \mbox{eff} }[g_{ab}]$
with respect to the background metric $g_{ab}$. 
Let us summarize
the principal feature of that method very briefly.\\

In a closed (namely, compact without boudary) Euclidean manifold,
the one-loop renormalized stress tensor of a scalar field $\phi(x)$ of
a $\zeta$-{\em regular theory}\footnote{That is a QFT theory
involving  meromorphic $x$-smooth  local $\zeta$ functions obtained 
by analytic continuation of  corresponding series, such that 
the function  $\zeta(s,x|A)$  is
analytic in $s=0$ and  $\zeta_{ab}(s,x|A)$
takes a simple pole at $s=1$.} (see \cite{Zmoretti})
is obtained as
\begin{eqnarray}
\langle T_{ab}(x)\rangle &=& \left\{
\zeta_{ab}(s+1,x|A) + \frac{1}{2}\: g_{ab}(x) \: \zeta(s,x|A) \right.
 +\nonumber \\
 & &\left. +  s\left[ \zeta'_{ab}(s+1,x|A) + \ln (\mu^2) \zeta_{ab}(s+1,x|A)
 \right]
\right\}_{s=0} \label{general}.
\end{eqnarray}
 $\zeta(s,x|A)$ is the usual local $\zeta$ function of the Euclidean
second order motion operator $A$, namely, the motion equations are given by
$A\phi = 0$.
The tensorial $\zeta$ function $\zeta_{ab}(s,x|A)$ is obtained through
the $s$ analytic continuation in the whole complex plane  of the series
(where from now on  $'$ indicates that possible null eigenvalues are omitted)
\begin{eqnarray}
\zeta_{ab}(s,x|A) := {\sum_n}^{'} \lambda_n^{-s}\:
T_{ab}[\phi^*_n,\phi_n](x) \label{zab},
\end{eqnarray}
$T_{ab}[\phi^*_n,\phi_n](x)$ being the stress tensor evaluated on the 
normal modes of $A$.\\
We have seen that the definition above is equivalent to the simpler one
\begin{eqnarray}
<T_{ab}(x)> =
 \zeta_{ab}(1,x|A) + \frac{1}{2}\: g_{ab}(x) \: \zeta(0,x|A)
\label{regular},
\end{eqnarray}
which can be used in the case of a {\em super} $\zeta$-{\em regular theory},
namely, when the function $\zeta_{ab}(s,x|A)$ is analytic also in $s=1$
where a simple pole appears in general.
Notice that, in the considered case, the renormalization scale $\mu$ as well as
the finite renormalization counterterms (\cite{Zmoretti})
disappear from the final result. 
We have seen that such $\mu$-dependent counterterms, whenever they exist,
have a simple form related to the heat kernel coefficients, are conserved
and depend locally on the geometry: they represent a finite renormalization
of the coefficients which stay on the geometrical side of customary 
generalized Einstein's equations.\\
We remark that the formulae  given above arise by an opportune
 interpretation of the 
formal identity
\beq
<T_{ab}(x)> = -\frac{2}{g^{1/2}}\: \frac{\delta S_{\scriptsize\mbox{eff}}
[g]}{\delta g^{ab}(x)}\nonumber
\eeq
when we have computed the effective action through the $\zeta$ function
\beq
S_{\scriptsize \mbox{eff} }[g_{ab}] = \frac{1}{2} \:\frac{d\:\:\:}{ds}|_{s=0}
\:\zeta(s|A/\mu^2), \label{first0}
\eeq
where $\zeta(s|A) = \int d^4x \sqrt{g(x)}\: \zeta(s,x|A)$.

In \cite{Zmoretti}, we have seen that, dealing with a scalar field
(in general massive)
coupled with the curvature  through the parameter $\xi$ as usually and
propagating into an Euclidean  
compact without boundary manifold, the function $\zeta_{ab}(s,x|A)$ is
meromorphic and its poles are related to the heat kernel coefficients 
in a simple fashion. Hence, we have a $\zeta$-regular theory.
 
We have also proven that the definition of one-loop renormalized stress tensor 
given in (\ref{general}), (\ref{regular}) produces a conserved stress tensor
provided the manifold were closed and the theory $\zeta$-regular.
Actually, one can simply prove that the hypotheses of a compact without
boundary manifold is not necessary. Indeed, on noncompact manifolds, 
one can assume (\ref{general}) by 
definition considering spectral measure and integrations instead of 
summations. Thus,
  by means of 
the same  proof of the stress-tensor conservation used in 
\cite{Zmoretti} with a few changes,  one  gets that, 
if the involved $\zeta$ functions are
correctly meromorphic defining a $\zeta$-regular theory, the left hand
side of (\ref{general}) (or (\ref{regular})) will be automatically conserved. 
Similarly, in the same hypotheses, one finds the anomalous trace of a 
massless conformally coupled field in terms of the local
$\zeta$ function evaluated at $s=0$  \cite{Zmoretti} (i.e. the
heat kernel expansion coefficient $a_2(x|A)$, provided the local
heat kernel expansion holds true in the considered case\footnote{The presence
 of a boundary involves
well-known changes in  the heat kernel expansion we shall not consider here.}).

In \cite{Zmoretti} we have also considered identities as
\begin{eqnarray}
-\frac{\partial \ln Z_\beta
}{\partial\beta} \:
= -\int_\Sigma d\vec{x} \:\sqrt{-g_L}\: \langle T_{L\: 0}^{\:\:0}(\vec{x})
\rangle_\beta
 \label{thermal}
\end{eqnarray}
and
\begin{eqnarray}
-\frac{\partial \ln Z_\beta}{\partial L_i} \:
= -\frac{\beta}{L_i}
\int_\Sigma d\vec{x} \:\sqrt{-g_L}\: \langle T_{L\: i}^{\:\:i}(\vec{x})
\rangle_\beta,
 \label{thermal2}
\end{eqnarray}
where we  defined $\ln Z_\beta := S_{\scriptsize \mbox{eff} }$,
provided the (Euclidean and Lorentzian)
manifold admit a
global  (Lorentzian time-like)
Killing vector arising from the Euclidean temporal coordinate with
a  period $\beta =1/T\:$.
 $\vec{x}$ represents the spatial coordinates which belong to the
spatial section $\Sigma$ and $g_L=-g$
is the determinant of the Lorentzian metric. 
Obviously,  $Z_\beta$ has to be interpreted  as a {\em partition function}.
 Notice that
all quantities which appear in the formula above do not depend
on the Euclidean or Lorentzian time  because  the manifold is stationary
and thus no time dependence arises from the metric. Furthermore
the analytic continuation to the Lorentzian time yields\footnote{Throughout
the paper we shall employ the signature $(1,1,1,1)$ for the Euclidean metric
and $(-1,1,1,1)$ for the Lorentzian one.}
 $\langle T_{L\: 0}^{\:\:0}(\vec{x})\rangle =\langle T_{0}^0(\vec{x})
\rangle$ and 
$\langle T_{L\: i}^{\:\:i}(\vec{x})\rangle =\langle T_{i}^i(\vec{x})
\rangle$. 
Eq.(\ref{thermal2}) holds
assuming both the homogeneity along $x^0$ and $x^{i}$, $L$ being the ``period''
of the manifold along $x^{i}$.\\
We have proved both identities above 
requiring explicitly a compact (finite) spatial section without boundary
 of the manifold.
It is not obvious whether or not such identities hold dropping that hypothesis
and introducing some cut-off in both sides of the considered equations
in order to handle finite quantities.\\
A further subtlety which arises dropping our simple hypotheses
is related to the ambiguity in defining the right hand side
of the couple of identities above. Indeed, let us consider the first
identity above for example.  From the statistical mechanics, one expects 
to find, in the right hand side,
 the averaged Hamiltonian instead of the spatial integral of 
the averaged $T_0^0$. Actually,
the latter differs from the former just by  total spatial derivative
of objects (built up by employing the field fluctuations 
$\langle \phi^2(x)\rangle_\beta$)
which vanish after one integrates provided the spatial 
manifold have
 no boundary or the field does vanish on this, whenever it exists.
This situation could change dramatically in the presence
 of boundaries  or infinite spatial sections and, in general, has
 to be analyzed case by case.\\

In \cite{phi} D. Iellici and the author of the present paper introduced 
also a method to renormalize the one-loop field fluctuations  
$\langle \phi^2(x) \rangle$
in a curved spacetime. Also that method is based on a local $\zeta$ function
and, in principle, holds when the involved Euclidean manifold is closed.
Nevertheless we expect that it holds in a more general case.\\
We have checked the method in several manifolds obtaining reasonable results.\\
The field fluctuations are regularized as
\beq
\langle
\phi^2(x)\rangle = \frac{d}{ds}|_{s=0} \frac{s}{\mu^2} \zeta(s+1,x|A/\mu^2) 
\label{phi}.
\eeq
Notice that, if the local $\zeta$ function of the effective action
$\zeta(s+1,x|A/\mu^2)$ is analytic at $s=1$, the formula 
above reduces to the more usual expression also expected from the
Green function analysis
\beq
\langle\phi^2(x)\rangle = \zeta(1,x|A)
 \label{phireg}.
\eeq
Notice the disappearance of the scale $\mu$.\\
The renormalized field fluctuations play an important role also in
the relationship between the renormalized stress tensor and the 
renormalized Hamiltonian \cite{FF,Zmoretti} as stressed above.
We shall come back on this point in the final discussion.\\

In this paper we shall analyze our method to renormalize the stress tensor
and field fluctuations
in three different situations. First we shall consider the (thermal) theory
of a conformally coupled massless scalar field 
within the closed  Einstein universe. The Euclidean related manifold 
satisfies completely our initial hypotheses of a closed manifold.\\
Secondly, we shall consider   the same field propagating in the open
Einstein universe. The related Euclidean manifold is not compact and
this is a first nontrivial ground where check our approach assumed by
definition. \\
The third case we shall consider  is the Euclidean manifold related
both to the cosmic string manifold and Rindler space 
(which can be considered also as the manifold containing a 
very large mass black hole). 
That Euclidean manifold 
is not {\em ultrastatic} differently from the two manifolds considered above,
moreover, it
has a conical singularity which, for some aspects,
 could be 
considered as a boundary.
That singularity involves a lot of difficulties dealing with $\zeta$
function approaches to renormalize the effective action. In particular,
  stress tensor components
 built up through the local $\zeta$ function of the effective action
in the physical manifold
have been obtained 
making direct use of mechanical-statistical laws 
 or supposing {\em  a priori} a particular 
form of the stress tensor. These results disagree, at law energies,
  with those obtained by the point-splitting method
 (see Section II of \cite{moiel} and
the final discussion  of \cite{moiel} for a  discussion and references 
on these topics).
In this paper we shall see that, as for it concerns the stress tensor
in the conical manifold,
 it is possible to get the same results
arising also from the point-splitting approach, for every
value of the coupling parameter $\xi$ by means of our local $\zeta$
 function approach.
This result will be carried out not  depending on the mechanical-statistical
laws and without  supposing any particular form of the stress tensor
{\em a priori}.
In the final discussion we shall perform some remark on the problem
of the choice  of the configuration-space measure in the path integral 
to define the partition function of the fields. This problem becomes
important as far as the physics in the conical manifold is concerned.
 We shall see that
these problems are related to the renormalization procedure involved
in defining physical quantities, concerning the Hamiltonian in particular. 
These problems have become important after the  interesting works
\cite{FFZ,FF,FF2} where the renormalized value of the vacuum fluctuations 
plays a
central role in explaining the finite Bekenstein-Hawking entropy
in the approach of the induced gravity.

\section{Einstein's closed static universe}

The ultrastatic metric of the (Euclidean) 
Einstein closed static universe is \cite{bd}
\begin{eqnarray}
ds^2_{ECS} = d\theta^2 + g_{ij}dx^idx^j = d\theta^2 + a^2 \left( 
dX^2 + \sin^2X d\Omega^2_2 \right). \nonumber
\end{eqnarray}
$X$ ranges from $0$ to $\pi$ and $d\Omega^2_2$ is the usual metric
on $S_2$. The time coordinate $\theta$ ranges from $0$ to $\beta \leq +\infty$.
$\beta$ is the inverse temperature of the considered thermal state
referred to the Killing vector generated by the Lorentzian time
$i\theta$. The related
vacuum state corresponds to the  choice $\beta = +\infty$.
The curvature of the space is $R= 6/a^2$ and the Ricci tensor reads $R_{ij} =
2g_{ij}/a^2$, the remaining components vanish.\\
This manifold is closed, namely compact without boundary. Also
the spatial section at $\theta=$ constant are closed and their  volume  is
$V= 2\pi^2 a^3$. \\
Let us consider a conformally coupled massless scalar
field propagating within this manifold. We want to compute its stress tensor
referred to the thermal states  pointed out above, in particular we want to
get the vacuum stress tensor which is known in literature
\cite{bd}. Notice that all the hypotheses required in 
\cite{Zmoretti} to implement the stress-tensor $\zeta$-function approach
are  fulfilled.\\
Let us build up the function $\zeta_{ab}(s,x|A)$ necessary to get
$\langle T_{ab}(x)\rangle_\beta$ through (\ref{general}) or (\ref{regular}).
The general expression of 
$\zeta_{ab}(s,x|A)$ 
is \cite{Zmoretti}
\beq
\zeta_{ab}(s,x|A) &=& \bar{\zeta}_{ab}(s,x|A) 
-\xi\nabla_a\nabla_b \zeta(s,x|A) + \left(\xi - \frac{1}{4}
\right)g_{ab}(x) \Delta \zeta(s,x|A) \nonumber\\
& & + \xi R_{ab}(x) \zeta(s,x|A)
-\frac{1}{2} g_{ab}(x) \zeta(s-1,x|A), \label{zzbar}
\eeq
where, in the sense of the analytic continuation of both sides in the
whole $s$ complex plane:
\beq
\bar{\zeta}_{ab}(s,x|A) = {\sum_k}^{'} \lambda_k^{-s} \: \nabla_a\phi^*_k(x)
\nabla_b\phi_k(x) \label{zbarab}.
\eeq
We are interested in the case $\xi=\xi_c:=1/6$ (conformal coupling 
in four dimensions).
The local $\zeta$ function is similarly given by
\beq
\zeta(s,x|A) = {\sum_k}^{'} \lambda_k^{-s} \: \phi^*_k(x)
\phi_k(x) \label{zeta}.
\eeq
The functions $\phi_k(x)$ define a normalized complete set of
eigenvectors of the Euclidean motion operator:
\beq
A \phi_k = \lambda_k \phi_k\nonumber,
\eeq
where, in our case
\beq
A = -\partial^2_\theta -a^{-2} \Delta_{S_3} + \xi_c R\nonumber,
\eeq
The explicit form of the considered eigenvalues and Kroneker's 
delta-normalized eigenvectors is
well-known \cite{bd}. In particular we have $k \equiv (n, q, l, m)$ where
$n = 0, \pm 1, \pm 2, \pm 3,...$, $q = 1, 2, 3,...$, $l = 0, 1, 2,..., q-1$,
 $m = 0, \pm 1,\pm 2,..., \pm l $ and 
\beq
\lambda_k = \left(\frac{2\pi n}{\beta}\right)^2 + \left(\frac{q}{a}\right)^2
\label{eigenvalues}.
\eeq
The following relations, which hold true for normalized eigenvectors, 
 are also useful. We leave the proofs of these to the reader.
\beq
\sum_{lm} \phi_k^*(x)\phi_k(x) = \frac{q^2}{V\beta}, \label{first}
\eeq
notice that the right hand side of the equation above 
is nothing but the  degeneracy of each eigenspace times $1/2\beta V$ (or
$1/\beta V$ when $n=0$);
\beq
\sum_{lm} \partial_i\phi_k^*(x)\partial_j\phi_k(x) = g_{ij}(x)
\frac{q^2(q^2-1)}{3
V\beta a^2}, \label{second}
\eeq
and ($x^0 := \theta$)
\beq
\sum_{lm} \partial_0
\phi_k^*(x)\partial_0\phi_k(x) = \frac{(2\pi n q)^2}{V\beta^3} \label{third}.
\eeq
We have also, because of the homogeneity of the space
\beq
\zeta(s,x|A) =\frac{ \zeta(s|A)}{V\beta}, \label{local}
\eeq
where $\zeta(s|A)$ is the global $\zeta$ function obtained
by summing over $\lambda^{-s}_k$ as usually
\beq
\zeta(s|A) = {\sum_k}^{'} \lambda_k^{-s} \label{zetag}.
\eeq
It is possible to relate the function $\bar{\zeta}_{ab}(s,x|A)$ to the function
$\zeta(s,x|A)$. Indeed, we notice that
\beq
\lambda_k^{-s} \left(\frac{2\pi n}{\beta}\right)^2 =
\frac{\beta}{2(s-1)} \frac{\partial \lambda_k^{-(s-1)}}{\partial \beta}.
\nonumber 
\eeq
The identity above inserted into the definition (\ref{zbarab}) for $a=b=0$,
taking (\ref{third}) into account, 
 yields
\beq
\bar{\zeta}_{00}(s+1,x|A) = \frac{1}{2Vs} \frac{\partial\:\:\:}{\partial\beta}
\zeta(s|A), \label{zbar00}
\eeq
or equivalently
\beq
\zeta_{00}(s+1,x|A) = 
\bar{\zeta}_{00}(s+1,x|A)
 = -\frac{a}{2V\beta s} \frac{\partial\:\:\:}{\partial a}
\zeta(s|A) + \frac{\zeta(s|A)}{V\beta}, \label{zbar00'}
\eeq
which follows from the  identity above taking account of
\beq
2s\zeta(s|A) = \beta \frac{\partial\:\:\:}{\partial \beta}
\zeta(s|A) + 
a \frac{\partial\:\:\:}{\partial a}
\zeta(s|A). \label{symmetry}
\eeq
The last identity is a simple consequence of the expression of the
eigenvalues (\ref{eigenvalues}).\\
Concerning the components $ij$ (the remaining components vanish)
we can take advantage from the identity
\beq
\lambda_k^{-s} q^2 = 
\frac{3 a^3}{2(s-1)} \frac{\partial \lambda^{-(s-1)}_k}{\partial a} \label{id}.
\eeq
Inserting this  into (\ref{zbarab}) for $a=i, b=j$, taking 
(\ref{second}) into account, it arises
\beq
\bar{\zeta}_{ij}(s+1,x|A) = \frac{g_{ij}(x)}{3V\beta a^2}
 \left[ 
-\zeta(s+1|A) + \frac{a^3}{2s} \frac{\partial\:\:\:}{
\partial a}\zeta(s|A)\right]. \label{zbarij}
\eeq
To get the renormalized stress tensor,
we have to compute $\zeta(s|A)$ or equivalently $\zeta(s,x|A)$ only.
The expansion  of the latter 
over the eigenvalues reads
\beq
\zeta(s,x|A) &=& 
\frac{2}{V\beta}
\sum_{q=1}^{+\infty}
\sum_{n=1}^{+\infty}
q^2\left[ \left(\frac{2\pi n}{\beta}\right)^2+ 
\left(\frac{q}{a}\right)^2\right]^{-s} 
+ \frac{1}{V\beta}\sum_{q=1}^{+\infty}
q^2 \left[ \left(\frac{q}{a}\right)^2  \right]^{-s} \nonumber\\
&=&
\frac{2}{V\beta}
\sum_{q=1}^{+\infty}
\sum_{n=1}^{+\infty}
q^2\left[ \left(\frac{2\pi n}{\beta}\right)^2+ 
\left(\frac{q}{a}\right)^2\right]^{-s} 
+ \frac{a^{2s}}{V\beta} \zeta_R(2s-2) \label{z1}.
\eeq
The last $\zeta$ function is  Riemann's one.\\
Let us introduce the Epstein function \cite{zerbinil}
 obtained by continuing (into a meromorphic function) 
the series in the variable $s$
\beq
E(s,x,y) := \sum_{n,m=1}^{+\infty} \left( x^2 n^2 + y^2 m^2 \right)^{-s}
\label{epstein}.
\eeq 
We get trivially
\beq
\sum_{n,m=1}^{+\infty} m^2\left( x^2 n^2 + y^2 m^2 \right)^{-s}
= -\frac{1}{2y(s-1)}\frac{\partial\:\:\:}{\partial y}
E(s-1,x,y).
 \nonumber
\eeq
Employing such an identity, we can rewrite the expression (\ref{z1}) of
$\zeta(s,x|A)$ as
\beq
\zeta(s,x|A)
= \frac{a^{2s}}{V\beta} \zeta_R(2s-2) + \frac{a^3}{V\beta (s-1)}
\frac{\partial\:\:\:}{\partial a}
E(s-1,\frac{2\pi}{\beta},\frac{1}{a}). \label{z1medio}
\eeq
no  expression of the Epstein function in terms of elementary functions 
 exists in
literature. Anymore, there exist a well-know expansion in terms of
MacDonald functions \cite{zerbinil}
\beq
E(s,x,y)
&=& -\frac{1}{2} y^{-2s}\zeta_R(2s)
+ \frac{\sqrt{\pi} \Gamma(s-1/2)}{2x\Gamma(s)}
y^{1-2s} \zeta_R(2s-1) \nonumber\\
& &
+\frac{2\sqrt{\pi} x^{-2s}}{\Gamma (s)}
\sum_{m,n=1}^{+\infty} \left(\frac{\pi x m}{y n}
\right)^{s-1/2} K_{s-1/2}\left( \frac{2\pi y n m}{x}\right) \label{expansion1}.
\eeq
Notice that, due to the negative exponential behaviour 
of MacDonalds functions
$K_a(x)$ at large arguments, the last series defines a function 
which is analytic on the whole $s$ complex plane. The structure of the poles
of the Epstein function is due to the gamma and (Riemann's) zeta functions
in the first line of the formula above.
In particular there are only two simple poles at $s=1/2$ and $s=1$.\\
Taking account of the expression above and (\ref{z1medio}), we find
\beq
\zeta(s,x|A) = \frac{\sqrt{\pi}}{4\pi V} \frac{\Gamma(s-3/2)}{\Gamma(s)}
(2s-3) a^{2s-1} \zeta_R(2s-3) 
-\frac{a}{V \Gamma(s)} \left(
\frac{\beta}{2\pi}\right)^{2s-2} \Xi(s, \beta/a), \label{z1fine} 
\eeq
where the function $\Xi(s,\beta/a)$
given by
\beq
\Xi(s,z)
= 2\pi \frac{d\:\:\:}{d z}
  \sum_{m,n=1}^{+\infty} \left(\frac{2\pi^2  m}{z n}
\right)^{s-3/2} K_{s-3/2}\left(n m z \right) \label{expansion},
\eeq
 is analytic throughout the s complex
plane and, due to the 
large argument behaviour
of the MacDonald functions,
 vanishes as $\beta \rightarrow +\infty $ like $(\beta/a)^{5/2-s}
\exp{-\beta/a}$ when Re $s \geq 0$.\\
Reminding the relation 
\beq
2\frac{d\:\:\:}{d u} K_a(u)
= K_{a-1}(u) +K_{a+1}(u) 
\eeq
the function $\Xi(s,z)$ and its $z$ derivative (see below) can be
evaluated numerically at the physical values $s=0$ and $s=1$
(see below).\\
The expression (\ref{z1fine}) which 
is very useful as far as the low temperature thermodynamics
in our manifold is concerned. Notice that, changing the role of 
$x$ an $y$ in the expression (\ref{expansion}), one may get an expression
for  $\zeta(s,x|A)$ useful at large temperatures.\\
Some remarks on (\ref{z1fine}) are in order. First notice that, 
due to the gamma functions into the denominators, $\zeta(s,x|A) \rightarrow 
 0$ like $s$ when $s\rightarrow 0$
and thus no trace anomaly appears and neither renormalization scale $\mu$
remains in the renormalized effective action. The found $\zeta$ function
is analytic throughout the $s$ complex plane except for the point $s=2$
where a simple pole appears.
Employing (\ref{zbar00}) and (\ref{zbarij}) we find that $\zeta_{ab}(s,x|A)$
is analytic at $s=1$ and thus the theory is a {\em super} $\zeta$-regular
theory.\\

Employing the definition (\ref{regular}), (\ref{zzbar})
 and the obtained expression for
$\zeta_{ab}(s,x|A)$, a few calculations lead us to
\beq
\langle T_{La}^{\:\:b}(x)
 \rangle_\beta = \langle T_{a}^b(x) \rangle_\beta \equiv 
T(\beta) \:
(-1, \frac{1}{3},\frac{1}{3},\frac{1}{3}) \label{finalclosed},
\eeq
where
\beq
T(\beta) =
-\frac{1}{2V} \frac{\partial\:\:\:}{\partial\beta} \frac{\zeta(s|A)}{s}|_{s=0}
&=& \frac{1}{480 a^4 \pi^2} 
+ \frac{1}{a^4} \frac{d\:\:\:}{dz}|_{z=\beta/a} \frac{\Xi(0, z)}{z}
\label{T}.
\eeq
Notice that the last  derivative term vanishes very fast
at low temperatures. \\
Now, one can  prove very simply  that the
 obtained stress tensor is conserved, has a vanishing trace and reduces to the
well-known  vacuum stress tensor in the closed Einstein universe \cite{bd}
as $\beta \rightarrow +\infty$
\beq
\langle T_a^b(x) \rangle_{\scriptsize \mbox{vacuum}}
 \equiv \frac{1}{480 a^4 \pi^2} \:
(-1, \frac{1}{3},\frac{1}{3},\frac{1}{3}) \label{finalclosedvacuum}.
\eeq
Taking account of $\zeta(0|A)=0$, we can rewrite (\ref{T}) as
\beq
T(\beta) =
-\frac{1}{2V} \frac{\partial\:\:\:}{\partial\beta} \zeta'(0|A)
= -\frac{1}{V} \ln Z_\beta \nonumber
\eeq 
where the prime means the $s$ derivative.
Hence, the relation (\ref{thermal}) holds true trivially.
The general relation between the Hamiltonian density and the stress-tensor
energy density in case of  static coordinates reads\footnote{Notice
that we are writing
 Lorentzian relations employing the Euclidean metric.
We could pass to use the  more usual Lorentzian 
metric simply through the identities
$g=-g_L$, $g_{00}= -g_{L00}$ and $g^{ij}= g^{ij}_L$.} \cite{FF}
\begin{eqnarray}
{\cal H} =   -T^0_0  + \xi g^{-1/2} \partial_i [g^{1/2}
(g^{ij} \partial_j \phi^2   \:- \phi^2 w^{i}) ] \label{TH},
\end{eqnarray}
  where $w^{a} = \frac{1}{2}
\nabla^{a} \ln g_{00}$. $w^{a}$ vanishes in the present case.
Let us employ such a relationship to evaluate the averaged value of the 
quantum Hamiltonian. We have to interpret (\ref{TH}) as
\begin{eqnarray}
\langle {\cal H} \rangle_\beta =   -
\langle T^0_0 \rangle_\beta + \xi g^{-1/2} \partial_i [g^{1/2}
(g^{ij} \partial_j \langle \phi^2 \rangle_\beta  \:- 
\langle \phi^2\rangle_\beta w^{i}) ] \label{TH'}.
\end{eqnarray}
As is well-known, provided the local $\zeta$ function were  regular
at $s=1$,  we can use the relation (\ref{phireg}) to evaluate 
$\langle \phi^2(x) \rangle$. This is the case and we find
\beq
\langle \phi^2(x) \rangle_\beta
 = -\frac{1}{48 \pi^2 a^2} - \frac{1}{2\pi^2 a^2}
\Xi(1,\beta/a). \nonumber
\eeq 
This reduces to the known value as $\beta \rightarrow +\infty$ \cite{bd}. 
Notice that, due to the homogeneity of the space, there is not dependence on
$x$ and thus all derivatives in (\ref{TH'}) vanish yielding
$\langle {\cal H}\rangle_\beta =   -\langle T^0_0\rangle_\beta$.
Then (\ref{thermal}) can be  rewritten in terms of the averaged
Hamiltonian in the right hand side
\begin{eqnarray}
-\frac{\partial \ln Z_\beta
}{\partial\beta} \:
=  \langle H \rangle_\beta
 \label{thermalH}
\end{eqnarray}

\section{Einstein's open static universe}

The ultrastatic metric of the (Euclidean)
Einstein closed static universe is \cite{bd}
\begin{eqnarray}
ds^2_{EOS} = d\theta^2 + g_{ij}dx^idx^j = d\theta^2 + a^2 \left(
dX^2 + \sinh^2X d\Omega^2_2 \right). \nonumber
\end{eqnarray}
$X$ ranges from $0$ to $+\infty$ and $d\Omega^2_2$ is the usual metric
on $S_2$. The time coordinate $\theta$ ranges from $0$ to $\beta \leq +\infty$.
Again, $\beta$ is the inverse temperature of the considered thermal state
referred to the Killing vector generated by the Lorentzian time
$i\theta$ and the  related
vacuum state corresponds to the  choice $\beta = +\infty$.
The curvature of the space is $R= -6/a^2$ and the Ricci tensor reads $R_{ij} =
-2g_{ij}/a^2$, the remaining components vanish.\\
This manifold is not closed and the spatial sections 
have not a finite volume.\\

Let us consider a conformally coupled massless scalar
field propagating within this manifold. As in the 
previously considered case, we want to compute its stress tensor
referred to the thermal states, in particular we want to
get the vacuum stress tensor. Notice that not all the hypotheses required in
\cite{Zmoretti} to implement the stress-tensor $\zeta$-function approach
are  fulfilled.   The manifold has no boundary but it is not compact.
We expect to find a continuous spectrum as far as the Euclidean 
motion operator is concerned.\\
However, we shall find that our method does work also in this case.
Notice that, now, we have to  employ our definition (\ref{general})
or (\ref{regular}) by definition and check on the obtained results
finally.

The form of the  eigenvalues $\lambda_k$ of the conformally coupled
massless Euclidean motion operator
\beq
A = -\partial^2_\theta -a^{-2}\Delta_{H_3} + \xi_c R\nonumber,
\eeq
 is
well-known \cite{bunch,bd}, 
we have, exactly as in the previous case 
\beq
\lambda_k = \left(\frac{2\pi n}{\beta}\right)^2 + \left(\frac{q}{a}\right)^2
\label{eigenvaluesO},
\eeq
where 
 $k \equiv (n, q, l, m)$ and
$n = 0, \pm 1, \pm 2, \pm 3,...$ , $q \in [0,+\infty)$,
$l= 0,1,2,3,...$
 $m = 0, \pm 1,\pm 2,..., \pm l $. The degeneracy depends only on
the indexes $l$ and $m$.\\
The following relations which hold true for  eigenvectors
$\phi_k(x)$ (which are Dirac's delta normalized in $q$ and Kroneker's delta
normalized in the remaining variables) 
 are also useful. We leave the proofs of these to the reader
(see also \cite{bunch}).
\beq
\sum_{l,m} \phi_k^*(x)\phi_k(x) = \frac{q^2}{2\pi^2 a^3 \beta}, \label{firstO}
\eeq
\beq
\sum_{l,m} \partial_i\phi_k^*(x)\partial_j\phi_k(x) = g_{ij}(x)
\frac{q^2(q^2+1)}{6\pi^2 a^5 \beta }, \label{secondO}
\eeq
and ($x^0 := \theta$)
\beq
\sum_{l,m} \partial_0
\phi_k^*(x)\partial_0\phi_k(x) = \frac{(2\pi n q)^2}{2\pi^2 a^3\beta^3} 
\label{thirdO}.
\eeq
Notice that the global $\zeta$ function simply does not exist because
the infinite spatial volume of the manifold.
Anyhow, we can compute the local $\zeta$ function as
\beq
\zeta(s,x|A) := \int_0^{+\infty}dq\: \sum_{l,m,n} \phi_k^*(x)
\phi_k(x) \lambda_k^{-s} \label{localO}.
\eeq
It is convenient to separate the contribution due to the terms with 
$n=0$ and introduce, as far as these terms are concerned, a cutoff
$\epsilon$ at low $q$. A few trivial manipulations of the expression
above yields
\beq
\zeta(s,x|A) = \frac{a^{2s-3}}{2\pi^2 \beta}
\int_\epsilon^{+\infty} dq \: q^{2s-2}
+ \frac{1}{4\pi^2 \beta} \left(\frac{\beta}{2\pi}\right)^{2s-3} \zeta_R(2s-3)
\frac{\Gamma(1/2) \Gamma(s-3/2)}{\Gamma(s)}  \label{local2O}.
\eeq 
The apparent divergent integral as $\epsilon \rightarrow 0^+$ can be made
harmless
as in \cite{hawking} putting 
$\epsilon \rightarrow 0^+$ after one has fixed Re $s$ large finite.
This procedure generalize 
the finite volume prescription to drop the null eigenvalues in defining
the $\zeta$ function for the case  of an infinite spatial volume.
We have finally
\beq
\zeta(s,x|A) = \frac{1}{8 \pi^2\sqrt{\pi} } \left( \frac{\beta}{2\pi}
\right)^{2s-4} \zeta_R(2s-3) \frac{\Gamma(s-3/2)}{\Gamma(s)} \label{localO3}.
\eeq
Notice that $\zeta(0,x|A) = 0$ and thus no renormalization scale 
appears in the (infinite) partition function.\\
Let us evaluate $\bar{\zeta}_{ab}(s,x|A)$. The only nonvanishing 
components are $00$ and $ij$. In the first case we have directly from the 
 definitions
(omitting the terms with $n=0$ as above)
\beq
\zeta_{00}(s+1,x|A) &= & \bar{\zeta}_{00}(s+1,x|A)
 =
\int dq \sum_{l,m,n} \left(\frac{2\pi n}{\beta}\right)^2
 \lambda_k^{-s} \phi_k^*(x)\phi_k(x)\nonumber\\
&=& \frac{1}{8\pi^2\sqrt{\pi}} \left( \frac{\beta}{2\pi}
\right)^{2s-4} \zeta_R(2s-4) \frac{\Gamma(s-1/2)}{\Gamma(s+1)}.
\label{zbar00O}
\eeq
In order to compute the remaining components of $\bar{\zeta}_{ab}$ we can use 
(\ref{secondO}) and the relation in (\ref{id}) once again.
We find
\beq
\bar{\zeta}_{ij}(s+1,x|A) = \frac{1}{3a^5} g_{ij}(x)
\zeta(s+1,x|A) + \frac{1}{2s} g_{ij}(x) \zeta(s,x|A). 
\eeq
We have found  that $\zeta_{ab}(s,x|A)$ is analytic in $s=1$, hence
the theory is a {\em super} $\zeta$-regular theory once again. We can use 
(\ref{regular}) to compute the stress tensor.\\
Through (\ref{zzbar}) and (\ref{regular}) we find finally
\beq
\langle T_{La}^{\:\:b} \rangle_\beta =\langle T_a^b \rangle_\beta 
\equiv
T(\beta) (-1,\frac{1}{3},\frac{1}{3},\frac{1}{3})  \label{tabO},
\eeq
where
\beq
T(\beta) = \frac{\pi^2}{30\beta^4} \label{tbetaO}.
\eeq
The stress tensor in (\ref{tabO}) is conserved and  traceless
as we expected from the general theory. $\langle T_a^b \rangle_\beta$
vanishes as $\beta \rightarrow +\infty$, this agrees with the known
result \cite{bd} that the stress tensor in the vacuum state of the
open Einstein universe vanishes.\\
Notice that the found stress tensor,
in the considered 
components, is exactly the same than in Minkowski spacetime.\\

Let us finally consider (\ref{thermal}). In this case the left hand side of 
(\ref{thermal})
does not exist because that simply diverges.
Nevertheless, we can notice that the divergence of the partition function
is due to the volume divergence only and 
the remaining factor does not depend on the position on the spatial section,
namely
\beq
\ln Z_\beta = V \ln {\cal Z}_\beta = V \times \left(\beta \frac{1}{2}
\zeta'(0,x|A)\right) = V \times \frac{\pi^3}{90\beta^3}  \label{xyz}
\eeq
where $V$ diverges and, actually, $\zeta'(0,x|A)$ 
does not depend on $x$ due to the
homogeneity of the spatial manifold. 
This is the same situation than arises in the Minkowski spacetime.\\
 We expect that,
although (\ref{thermal}) does not make sense, a local version could yet
make sense. Indeed, one can get very simply from (\ref{tabO}) 
and (\ref{xyz}) 
\begin{eqnarray}
-\frac{\partial \:V \ln {\cal Z}_\beta}{\partial\beta} 
= - V \: \langle T_{0}^{0}\rangle_\beta =
 -\int_V d\vec{x} \:\sqrt{g}\: \langle T_{L\: 0}^{\:\:0}(\vec{x})
\rangle_\beta
\eeq
on any finite volume $V$.
 As in the previously discussed case,
$\langle \phi^2(x) \rangle_\beta$ 
can be obtained  through (\ref{phireg}) \cite{phi}
\beq
\langle \phi^2(x) \rangle_\beta = \frac{1}{12\beta^2}.
\eeq
Notice that this vanishes as $\beta \rightarrow +\infty$
namely, in the vacuum state as is known \cite{bunch}. Furthermore, it 
does not depend on $x$ and thus, through (\ref{TH}) and noticing that $w^{a}
=0$
(see the Einstein closed universe case), 
$\langle T_0^0 \rangle_\beta = -\langle {\cal H} \rangle_\beta $.
We can write finally, with an obvious meaning
\begin{eqnarray}
-\frac{\partial \:V \ln {\cal Z}_\beta
}{\partial\beta} \:
=  \langle H_V \rangle_\beta.
 \label{thermalH2}
\eeq

\section{The conical manifold}

Let us consider the Euclidean manifold ${\cal C}_\beta \times \R^2$
 endowed with the metric
\beq
ds^2 = r^2 d\theta^2 + dr^2 + dz_1^2 + dz_2^2 \label{conic}, 
\eeq
where $(z_1,z_2)\in \R^2$, $r\in [0,+\infty)$, $\theta \in [0,\beta)$
when $0$ is identified with $\beta$. ${\cal C}_\beta \times \R^2$
is a cone with deficit angle given by $2\pi - \beta$.
That is the Euclidean manifold corresponding to the finite temperature
($T= 1/\beta$) quantum field theory in the Rindler space. In such a case
$\theta$ is the Euclidean time of the theory.
This is also a good
approximation of a large mass black hole near the event horizon.
Equivalently, considering $z_1$ as the Euclidean time, the metric above 
defines the Euclidean section (at zero temperature) of a cosmic string
 background. In this case $(2\pi-\beta)/8\pi G$ is the mass of the string.\\
The metric in (\ref{conic}), considered as the Rindler Euclidean metric,
 is static but not {\em ultrastatic}. Another important point is that
such a metric is not homogeneous in the spatial section.\\
The considered manifold is flat everywhere except for  
 conical singularities which appear at $r=0$ whenever $\beta \neq 2\pi$.
These singularities produces well-known 
Dirac's delta singularities in the curvatures
of the manifolds at $r=0$ \cite{singular}. The physics involved in such
anomalous curvature it is not completely known. Actually, we shall see 
shortly that
one can ignore completely the anomalous curvature dealing with the 
stress tensor renormalization also considering nonminimal coupling with
the scalar curvature.\\
As is well-known, the particular value $\beta_H = 2\pi$ defines
 the Hawking-Unruh temperature in the Rindler/large-mass-black-hole
interpretation, the corresponding thermal state being nothing but the
Minkowski vacuum/Hartle-Hawking (large mass) vacuum.\\
The thermal Rindler stress tensor (renormalized with respect to the 
Minkowski vacuum)
 which coincide, in the Euclidean approach, to
  the zero-temperature cosmic-string stress
tensor (renormalized with respect to the Minkowski vacuum) has been
computed by the point splitting approach \cite{frolov}.\\
Such results has been only partially reproduced by some
 $\zeta$-function or (local) heat kernel
approach \cite{kcv,ielmo}. This is because these approaches were employed to
renormalize the effective action only, and thus the stress tensor
was computed assuming further hypotheses on its form or assuming 
some statistical-mechanical law as holding true \cite{kcv,ielmo}.\\
 Recently, in \cite{iellici}, also
 the massive case has been considered by employing an off-diagonal 
$\zeta$-function approach and a subtraction procedure similar to
that is employed within  the point-splitting framework.\\
Here, we shall consider the massless case only. We shall check our approach
for every value of the curvature coupling proving that the same results
got by the point-splitting approach naturally arise.
The important point is that, due to the complete independence of the
method from statistical mechanics, we shall be able to discuss
 the statistical mechanics meaning (if it exists)
 of our results {\em a posteriori}.\\

Let us consider first the case of the {\em minimal} coupling $\xi=0$.
This avoids all problems involved dealing with the singular curvature
on the tip of the cone generated by the conical singularity.
The  
$\zeta$ function of the effective action  in  conic 
backgrounds 
 has been computed by several authors \cite{zcv} 
 also in the massive
scalar case \cite{iellici}
and for photons and gravitons \cite{moiel}. \\
Discarding the singular curvature by posing 
$\xi=0$, a
 complete normalized  set of  eigenvectors of the massless Euclidean motion
 operator\footnote{We are considering a particular 
self-adjoint extension of the formally
self-adjoint Laplace-Beltrami operator in the conical manifold.
The general theory of these extensions has been studied in \cite{kay}.}
 $A = -\Delta_{{\cal C}_\beta \times \R^2}$ 
 is \cite{zcv} 
\beq
\phi_q(x) = \frac{1}{2\pi} \sqrt{\frac{\lambda}{\beta}}\:
e^{ikz} e^{i\frac{2\pi n}{\beta} \theta}
J_{\frac{2\pi |n|}{\beta}}(\lambda r) \label{eigenvector3}
\eeq
where $z=(z_1,z_2)$,
 $q = (n,k,\lambda)$, $n= 0, \pm 1,\pm 2,...$, $k= (k_1,k_2)\in \R^2$
$\lambda \in [0, +\infty)$.
The considered eigenfunctions are Kroneker's delta normalized in the index
$n$ and Dirac's delta normalized in the remaining indices.
The corresponding eigenvalues are
\beq
\lambda_q = \lambda^2 + k^2  \label{eigenvalues3}.
\eeq
The $\zeta$ function of $A$ has been computed explicitly and reads
\beq
\zeta(s,x| A) = \frac{r^{2s-4}}{4\pi \beta \Gamma(s)} I_\beta(s-1)
\label{zeta3}.
\eeq
$I_\beta(s)$ is a well-known meromorphic function \cite{zcv}
carrying a simple pole at $s=1$. Known values are also
\beq
I_\beta(0) &=& \frac{1}{6\nu}(\nu^2 - 1) \label{I0},\\
I_\beta(-1)&=& \frac{1}{90 \nu}(\nu^2 -1)(\nu^2 +11) \label{I-1},
\eeq 
where we defined $\nu := 2\pi/\beta$.\\
Notice that $\zeta(0,x|A)=0$ and thus no scale remains into the renormalized
local effective action and $\langle \phi^2(x) \rangle$ can be computed through
(\ref{phireg}).\\
The function $\bar{\zeta}_{ab}(s,x|A)$
 can be computed making use of intermediate 
results contained in \cite{iellici}. A few calculations lead us to
\beq
\bar{\zeta}_{\theta\theta}(s,x|A) &=&  
\frac{r^{2s-4}\Gamma(s-3/2)}{4\pi \sqrt{\pi} \beta \Gamma(s)} H_\beta(s-1)
\label{thetatheta3},\\
\bar{\zeta}_{rr}(s,x|A) &=&  
\frac{1}{2r}\partial_r r\partial_r \zeta(s,x|A)
 - \frac{1}{r^2}
\bar{\zeta}_{\theta\theta}(s,x|A) + 4\pi (s-2) \zeta_{D=6}(s,x|A),
 \label{rr3}\\
\bar{\zeta}_{z_1z_1}(s,x|A) &=&  \bar{\zeta}_{z_2z_2}(s,x|A) = 
2\pi \zeta_{D=6}(s,x|A). \label{zz3}
\eeq
All remaining components vanish.
The meromorphic function $H_\beta(s)$ has been defined in \cite{iellici},
it has a simple pole at $s=2$ and known values are
\beq
H_\beta(0) &=& \frac{1}{120\nu}(\nu^4 - 1) \label{H0},\\
H_\beta(1)&=& -\frac{1}{12 \nu}(\nu^2 -1) \label{H1}.
\eeq 
The function $\zeta_{D=6}(s,x|A)$ is the $\zeta$ function of the 
effective action in ${\cal C}_\beta\times \R^4$ \cite{iellici}, it reads
\beq
\zeta_{D=6}(s,x| A) = \frac{r^{2s-6}}{(4\pi)^2 \beta \Gamma(s)} I_\beta(s-2)
\label{zeta6}.
\eeq
From the above equations and (\ref{zzbar}) it follows that
$\bar{\zeta}_{ab}(s,x|A)$ is analytic at $s=1$ and thus
 the theory is 
super $\zeta$-regular once again. Hence, we can use (\ref{regular}) to
compute the stress tensor.  Trivial calculations employing (\ref{zzbar})
with $\xi=0$ and (\ref{regular}) produce
\beq
\langle T_a^b(x)_{\xi=0} \rangle_\beta  &\equiv& 
\frac{1}{1440\pi^2 r^4}
\left\{ \left[ \left(\frac{2\pi}{\beta} \right)^4 -1 
\right]\mbox{diag}(-3,1,1,1)\right.\nonumber\\
& &- \left.
20\left[ \left(\frac{2\pi}{\beta} \right)^2 -1 
\right]\mbox{diag}(\frac{3}{2},-\frac{1}{2},1,1)
\right\} \label{xizero}.
\eeq
This is the correct result arising by the point-splitting approach
\cite{frolov} in the case of the minimal coupling. Let us prove that our method
reproduce also the remaining cases. \\
In general, the relationship  between  the minimally coupled stress-tensor
and the generally coupled stress tensor can be trivially obtained 
by varying the action containing the usual coupling with the curvature,
it  reads
\beq
T_{ab}(x)_\xi = T_{ab}(x)_{\xi=0} + \xi 
\left[ (R_{ab}- \frac{1}{2}g_{ab}R) \phi^2(x) + g_{ab} \Delta \phi^2(x) 
- \nabla_a \nabla_b \phi^2(x) \right] \label{class}.
\eeq
It is worthwhile stressing that the last $\xi$-parametrized term appears also
when the manifold is flat.
We can interpret quantistically  this relationship as
\beq
\langle T_{ab}(x)_\xi\rangle  = \langle T_{ab}(x)_{\xi=0}\rangle + \xi
\langle Q(x)_{ab}\rangle, \nonumber
\eeq
where
\beq 
\langle Q(x)_{ab}\rangle
 := \left[ (R_{ab}(x)- \frac{1}{2}g_{ab}(x)R(x)) \langle \phi^2(x)
\rangle + g_{ab} \Delta \langle \phi^2(x) \rangle 
- \nabla_a \nabla_b \langle \phi^2(x)\rangle \right] \label{Q}.
\eeq
Now 
$\langle T_{ab}(x)_{\xi=0}\rangle_\beta$
 is known by (\ref{xizero}), $R_{ab}(x)=0$, $R(x)=0$ and thus we can 
compute $\langle T_{ab}(x)_{\xi}\rangle_\beta$
employing the known value of 
$\langle \phi^2(x)\rangle_\beta$. We have,
through  (\ref{zeta3}) and 
(\ref{phireg}) 
\beq
\langle \phi^2(x)
\rangle_\beta = \frac{1}{48\pi^2 r^2}\left[\left(\frac{2\pi}{\beta}
\right)^2-1 \right] \label{phi3}.
\eeq
The final result is exactly that of the point-splitting approach: 
\beq
\langle T_{La}^{\:\:b}(x)_{\xi} \rangle_\beta  =
\langle T_a^b(x)_{\xi} \rangle_\beta  &\equiv& 
\frac{1}{1440\pi^2 r^4}
\left\{ \left[ \left(\frac{2\pi}{\beta} \right)^4 -1 
\right]\mbox{diag}(-3,1,1,1)\right.\nonumber\\
& &+ \left.
20(6\xi -1)\left[ \left(\frac{2\pi}{\beta} \right)^2 -1 
\right]\mbox{diag}(\frac{3}{2},-\frac{1}{2},1,1)
\right\} \label{xigen}.
\eeq
The same result arises by employing the definition of
 $\zeta_{ab}(s,x|A)$ given in (\ref{zzbar}) with the chosen value of 
$\xi$, provided $\bar{\zeta}_{ab}(s,x|A)$
and $\zeta(s,x|A)$ were those computed in the {\em minimal coupling case}.
This means that, concerning the renormalization of the stress tensor,
the presence of the conical singularity which determines
a singular curvature on the tip of the cone  is completely irrelevant.
Concerning the quantum state, there is no difference between different
 couplings with the curvature. The 
$\xi$-parametrized  
term remains as a relic 
in the stress tensor because of the {\em classical} formula (\ref{class}).
This term does not come out
 from the quantum state once one
fixed the renormalization procedure.  
We see that the renormalization of the stress tensor can be managed completely
 by our Euclidean $\zeta$-function
 approach on the physical manifold instead of the {\em optical} manifold 
(see the final discussion) 
 not depending on the presence of the conical singularity
in the Euclidean manifold.\\
The knowledge of the averaged and renormalized stress tensor makes us able 
to compute the averaged and renormalized Hamiltonian of the system. 
 The Hamiltonian of the theory should not depend on the parameter $\xi$
because that cannot appear into the Lorentzian action the manifold
being flat.  Notice that 
there is no conical singularity in the Lorentzian
 theory! 
Not depending on $\xi$, 
the classical Hamiltonian density coincides with the changed sign  
energy component of the
stress tensor in the minimal coupling.
Indeed, employing (\ref{TH}), we can write down
\begin{eqnarray}
 \langle {\cal H}(x) \rangle_\beta   
= -\langle T^0_0(x)_{\xi=0}\rangle_\beta = 
\frac{3}{1440\pi^2 r^4}
\left[  \left(\frac{2\pi}{\beta} \right)^4 
 - 10 \left(\frac{2\pi}{\beta} \right)^2 -11  \right] \label{xizero1}.
\end{eqnarray}\\
Let us finally 
consider the problem of  the validity of the relation (\ref{thermal})
 in some sense.
The spatial section  is neither finite nor homogeneous, we could have 
problems with the use of cutoffs. It is not obvious that such a relation as
(\ref{thermal}) can hold true in our case considering cutoff smeared quantities
as\footnote{Notice that also the area $A$ of the horizon is a cutoff because
the actual area is infinite. This cutoff is a trivial overall factor.
We shall omit this cutoff as an index in the 
following formulae for sake of simplicity.}
\beq
\ln Z_{\beta \epsilon} 
&:=& \int_{r>\epsilon} d^4x \sqrt{g}
\:\frac{1}{2}\frac{d\:\:\:}{ds}|_{s=0}\zeta(s,x|A), \label{zed}\\
{\cal Q}_{\epsilon}(\beta)
&:=&  \int_{r>\epsilon} d^3x \sqrt{g} 
\langle Q_0^0(x)\rangle_\beta,\\
\langle H_\epsilon \rangle_\beta &:=& 
\int_{r>\epsilon} d^3x \sqrt{g} 
\langle {\cal H}\rangle_\beta
\eeq
and finally 
\beq
{\cal E}_{\epsilon\xi}(\beta):=
  -\int_{r>\epsilon} d^3x \sqrt{g} 
\langle T_0^0(x)_\xi\rangle_\beta
 =  \int_{r>\epsilon} d^3x \sqrt{g} 
\langle {\cal H}\rangle_\beta
 - \xi {\cal Q}_{\epsilon}(\beta).
\eeq
In particular we have from (\ref{zeta3})
\beq
\ln Z_{\beta \epsilon} = \frac{A\beta}{2880 \pi^2 \epsilon^2}
\left[\left(\frac{2\pi}{\beta}\right)^4
 + 10 \left(\frac{2\pi}{\beta}\right)^2 -11 \right],
\label{abc}
\eeq
where $A$ is the area of the event horizon, the regularized volume 
of the spatial section is $V_\epsilon = A/(2\epsilon^2)$.
Notice that, actually, the conserved charge ${\cal Q}_\epsilon(\beta)$
is a boundary integral which diverges on the conical singularity.
Indeed, it can be expressed by the integration of (\ref{TH}) and it should
 discarded if the manifold were regular.
Notice that the choice of values of  $\xi$ determines  different values
of ${\cal E}_{\xi\epsilon}$  due to the $\xi$-parametrized
boundary term $\xi {\cal Q}_\epsilon$
 in the stress tensor. Conversely, $\ln Z_{\beta\epsilon}$
does not depend on $\xi$.\\
 If something like (\ref{thermal})
holds true for a fixed value of $\epsilon$,
 it does just for a particular and unique value of $\xi$. Actually 
few calculation through  (\ref{xigen})
prove that, {\em not  depending on the value of} $\epsilon$
\beq
- \frac{\partial \ln Z_{\beta \epsilon}}{\partial\beta} 
= {\cal E}_{\epsilon \xi=1/9}(\beta) + {\cal E}_{\epsilon} =
 \langle H_\epsilon \rangle_\beta - \frac{1}{9}{\cal Q}_\epsilon(\beta)
+ {\cal E}_\epsilon.
\eeq
The last term is an opportune
 constant energy
\beq
{\cal E}_{\epsilon} = \frac{A}{120 \pi^2 \epsilon^2}. \nonumber
\eeq
The presence of such an added constant
could be expected from the fact that  the energy
${\cal E}_{\epsilon\xi}$ is renormalized to vanish at $\beta=2\pi$ instead of 
$\beta= +\infty$. Conversely, there 
is no trivial explanation of the presence of
the $\beta$-{\em dependent}  term $- \frac{1}{9}{\cal Q}_\epsilon(\beta)$.
Then, in the considered case,
in the right hand side of (\ref{thermal}) 
does not appear the Hamiltonian which, at least 
classically, corresponds to the value $\xi=0$ as discussed above.\\
One could wonder whether or not $Z_{\beta \epsilon}$
defined in  (\ref{zed}) can be considered a (regularized) partition function
of the system. The simplest answer is obviously not because a fundamental
relationship of statistical mechanics does not hold true.\\
In general, one could think that this negative result 
arises because we have dropped a contribution due to the conical singularity.
This singularity produces a Dirac delta in the curvature on the tip of the
cone in the Euclidean manifold. The integral of the Lagrangian get a 
contribution from this term in the case of a nonminimal coupling with the
curvature. 
The problem of the contributions of these possible terms, in particular
in relation to the black-hole entropy 
has been studied by several authors
(see \cite{BS,FFZ,FF,H,M} and references therein), anyhow,
in this paper we shall not explore such a possibility.\\
In any cases, it is worthwhile stressing that
 the found Euclidean effective action (\ref{abc})
is the correct one in order to get the {\em thermal}
 renormalized stress tensor by (formal)
variation with respect to the background metric.
We re-stress that the obtained stress tensor is exactly that obtained by the 
point-splitting approach.

\section{Summary and discussion}

In this paper we have checked the method to renormalize the one-loop 
stress tensor
introduced in \cite{Zmoretti}. We have studied the application of the method 
to three different cases both considering the thermal
and non thermal QFT. The first case (the closed static Einstein universe)
fulfills completely the hypotheses requested in \cite{Zmoretti}. The found
results coincide with known results. The remaining two cases
(open Einstein static universe and Euclidean Rindler/cosmic-string
manifold) have concerned
two situations where the hypotheses requested in \cite{Zmoretti} were
not completely fulfilled. In particular the last case, the conical manifold,
could seem quite subtle concerning  $\zeta$ function techniques due to the
presence of the singular curvature. Actually, we have found that the 
presence of the conical singularity does not involves particular problems in
renormalizing the stress tensor because the kernel of the physical information
is completely contained in the minimally coupled case.
Anyhow, we have found correct results in both cases.
In particular, these results agree with those obtained by the point-splitting
approach.\\
In general, we expect that the method introduced in \cite{Zmoretti}
to renormalize the one-loop stress tensor could work
also relaxing the hypotheses of an Euclidean  closed manifold
as we have found in these examples.
Moreover, we expect that the results should coincide with those obtained
by the point splitting approach.\\

We shall conclude this work with some comments on the result found 
in the conical manifold.\\ 

There are still  some unresolved problems concerning the theory on the 
conical manifold interpreted as the thermal Rindler space.\\
The question whether or not the effective action computed by the $\zeta$ 
function defines also  the logarithm of the partition function 
(renormalized with respect to the Minkowski vacuum)
is not a simple question. \\
The problem is interesting on a physical ground 
also because the partition function of the field around a black hole
(we remind the reader that the Rindler metric represents a large mass
balck hole) is used to compute the quantum corrections to the
Bekenstein-Hawking entropy as early suggested by 't Hooft
\cite{hooft}
or  to explane the complete B-H entropy in the 
framework of the induced gravity considering massive fields
nonconformally coupled \cite{FFZ,FF,FF2}. \\
On a more general ground, the considered  problem is also interesting
because
there exist two not completely equivalent approaches
 to implement the statistical mechanics 
of a quantum field in a curved spacetime through the use of a path
integral techniques and, up the knowledge of the author,
there is not a definitive choice of the method.
In this work, we have employed the path integral
in the physical manifold instead of in the {\em optical} related manifold.
We remind the reader that in the case of a {\em static}
 but not {\em ultrastatic} 
spacetime, the naive approach based on the phase-space path
integral leads one to a definition of the partition function as an Euclidean
path integral performed in the configuration space within the
{\em optical}
 manifold\footnote{This is the ultrastatic manifold
conformally related to the physical manifold
by defining the optical metric through
$\tilde{g_{ab}} := g_{ab}/g_{00}$. The Euclidean action employed on the
optical manifold is the physical action conformally transformed (including
the matter fields)
following the conformal transformation written above.} 
instead of the physical one \cite{optical}. Other approaches
\cite{toms} lead one to the definition of the partition function as a path 
integral in the physical manifold. \\
When the spatial section of the space is 
regular (e.g. closed) and thus the 
path integral regularized through  the $\zeta$-function approach
yields a finite result, formal manipulations of the path integral 
prove that   these two different definitions lead to the same result
up to the renormalization of the zero point energy  \cite{report}.
 In such a case 
 these definitions are substantially 
equivalent.
 When the manifold is not regular, e.g. it has spatial
sections with an infinite volume or has boundaries, in principle one may
loose   such an equivalence.
Indeed,
as far as the effective actions are concerned  in our case  we have 
\beq
\ln Z_{\beta \epsilon}
 = \frac{A\beta}{2880 \pi^2 \epsilon^2}
\left[\left(\frac{2\pi}{\beta}\right)^4
 + 10 \left(\frac{2\pi}{\beta}\right)^2 -11 \right],
\nonumber
\eeq
and 
\beq
\ln Z^{\scriptsize \mbox{opt}}_{\beta \epsilon} =
 \frac{A\beta}{2880 \pi^2 \epsilon^2}
\left(\frac{2\pi}{\beta}\right)^4
\label{abcsopt}
\eeq
The latter result can be directly obtained  noticing that the optical
manifold of the Rindler space is the open Einstein
static universe \cite{bd}. Hence the latter effective action above is nothing
but that computed previously
in the open Einstein universe (in the conformal coupling).\\

Further comments on the renormalization of the Hamiltonian in the
conical manifold  are in order.\\
Considering the  effective action computed as a path integral in the optical
manifold we have
\beq
- \frac{\partial 
\ln Z^{\scriptsize \mbox{opt}}_{\beta \epsilon}}{\partial\beta} 
= {\cal E}_{\epsilon\xi=1/6}(\beta) + {\cal E}_\epsilon'           
\label{xyzt}
\eeq
 One could conclude that,  once again, 
there is not the Hamiltonian in 
the right hand side, also discarding the constant energy. Actually,
 this result involves more subtle considerations. Indeed, we shall prove that
this naive conclusion is not correct. \\
Let us suppose to implement the canonical QFT \cite{bd}
for a massless field conformally coupled
 directly on the optical manifold,
namely in the open Einstein static universe as it were the physical manifold. 
Obviously, we should
get exactly the effective action which appears in (\ref{xyzt}).
Furthermore,
  (\ref{xyzt}) is nothing but (\ref{thermalH}) and the right hand side of
(\ref{xyzt}) is nothing but the averaged $\epsilon$-regularized  
Hamiltonian of the QFT in the open Einstein universe.
 Such a Hamiltonian can be also
obtained as a thermal average of the Hamiltonian operator got from the
 canonical QFT employing the {\em normal order prescription}.\\ 
Implementing  the canonical quantization in the Rindler space for a
 massless scalar field, one trivially
finds that  an isomorphism exists between the Fock space built up on the
Fulling-Rindler vacuum and the Fock space built up on the natural vacuum of the
QFT in the open Einstein static universe (in the conformal coupling). Indeed, 
this isomorphism arises from the conformal relationship between
the wavefunctions of the particles related to the quantized 
fields. This relation defines a one-to-one map from the one-particle Hilbert
space of the Einstein open universe  to the one-particle Hilbert space
of the Rindler space which maintains the value 
of the corresponding indefinite  scalar products \cite{bd}.
 This map defines a unitary
isomorphism between the two Fock spaces provided one require that this
isomorphism transform the vacuum state of the Einstein open universe into
the Fulling-Rindler vacuum.
In particular, also the Hamiltonian operators are unitarily identified
provided one use the normal order prescription in both cases.\\
As a result we find that
 the right hand side of (\ref{xyzt}) coincides  also  with the averaged
Hamiltonian operator built up in the framework of the canonical quantization
in the Rindler space with respect to the Fulling-Rindler vacuum!\\
In this sense (\ref{xyzt}) is the usual statistical mechanical relationship
between the canonical energy and the partition function in the Rindler 
space.\\
The central point is that the renormalization scheme employed is the
normal order prescription with respect to the Fulling-Rindler vacuum and
not the point-splitting procedure.\\ 
We can finally 
compare the averaged Rindler Hamiltonian of the canonical quantization 
$\langle H_\epsilon^{\scriptsize \mbox{can}} 
\rangle_\beta$ which is renormalized by the {\em normal order prescription}
in the the Fulling-Rindler vacuum 
 with the averaged Rindler Hamiltonian $\langle H_\epsilon \rangle_\beta$
obtained by integrating (\ref{xizero1}).
The latter  is renormalized with respect the
 Minkowski 
vacuum  by the {\em point-splitting procedure}. 
 We find
\beq 
\langle H_\epsilon \rangle_\beta - \langle 
H_\epsilon^{\scriptsize \mbox{can}}  \rangle_\beta =
-\frac{3}{2880\pi^2 \epsilon^2} -\frac{30}{2880\pi^2 \epsilon^2} 
\left[ \left(\frac{2\pi}{\beta} \right)^2 -1 
\right] = -\frac{1}{960 \pi^2 \epsilon^2} - \frac{1}{6}
 {\cal Q}_\epsilon(\beta). \label{eqfine}
\eeq
The first term in the right hand side is trivial: it takes account of the
difference of the zero-point energy. The second term is quite unexpected.
It proves that the point-splitting procedure 
 (or equivalently our $\zeta$-function procedure)
to renormalize the stress
tensor and hence the Hamiltonian
is not so trivial as one could expect, this is 
 because it involves terms which do not represent a trivial
zero-point energy renormalization.\\

Concerning the conical manifold, the conclusion is that the theory 
in the optical manifold leads us naturally 
to an effective action which can be considered the logarithm
of the partition function provided we renormalize the theory with respect 
to the Fulling-Rindler vacuum.
 Conversely, the effective action evaluated in the 
physical manifold is the correct effective action which produces
the thermal stress tensor
by formal variations with respect to the metric (this is the method
exposed in \cite{Zmoretti}). This stress tensor
is that obtained also by the point-splitting procedure and thus
renormalizing with respect to the Minkowski vacuum.

\par \section*{Acknowledgments}

This work has been financially
 supported by the ECT* (European Centre for Theoretical
 Nuclear Physics and Related Areas).\\
I would like to thank R. Balbinot, D.V. Fursaev, D. Iellici, L. Vanzo
and A. I. Zelnikov for valuable discussions  and suggestions.\\
I am grateful to the Dept. of Physics of the Trento University
for the hospitality during a part of the  time spent to produce this paper.

\end{document}